\documentclass[preprint]{aastex}

\usepackage{amsmath}

\newcommand{\hs}{\,-\,}

\shorttitle{Radio wave emissions}
\shortauthors{Marklund, Brodin, Dunsby}

\begin{document}

\title{Radio wave emissions due to gravitational radiation}

\author{Mattias Marklund\altaffilmark{1,3,6}, Gert
Brodin\altaffilmark{2,4},
  and Peter K.\ S.\ Dunsby\altaffilmark{1,5}}

\altaffiltext{1}{Department of Mathematics and Applied Mathematics,
  University of Cape Town, Rondebosch 7701, South Africa}
\altaffiltext{2}{Department of Plasma Physics,
  Ume{\aa} University, SE--901 87 Ume{\aa}, Sweden}
\altaffiltext{3}{E-mail address: mattias.marklund@sto.foa.se}
\altaffiltext{4}{E-mail address: gert.brodin@physics.umu.se}
\altaffiltext{5}{E-mail address: peter@vishnu.mth.uct.ac.za}
\altaffiltext{6}{Present position: National Defence Research
  Establishment FOA, SE--172 90 Stockholm, Sweden}

%\date{\today}

\begin{abstract}
  We consider the interaction of a weak
  gravitational wave with electromagnetic fields in a thin plasma
  on a Minkowski background spacetime using the 1+3 orthonormal
  frame formalism. Because gravitational and electromagnetic
  waves satisfy the same dispersion relation, electromagnetic
  waves can be effectively generated as a result of this
  interaction. In the case of the interaction with a static
  magnetic field, the amplitude of the electromagnetic waves
  depends on the size of the excitation region in which the
  magnetic field is contained. It is argued that due to the
  presence of a plasma this process can also lead to the generation
  of higher harmonics of the original mode. Estimates are given
  for this effect in the case of a binary pulsar and a cold electron
  plasma. It is found that the emmited radiation will lie in the
  radio frequency band. We also speculate on the possible relevance
  of this process on situations in cosmology, in particular whether
  this could be used to constrain primordial magnetic fields.
\end{abstract}

%  \textit{PACS numbers:} 04.30.Nk, 52.35. g, 52.35.Mw, 95.30.Qd

%%%%%%%%%%%%%%%%%%%%%%%%%%%%%%%%%%
\section{Introduction}
%%%%%%%%%%%%%%%%%%%%%%%%%%%%%%%%%%
It is well known that electromagnetic (EM) waves can scatter off
gravitational fields. For instance, \citet{dewitt} considered raditation
damping in a gravitational field, and \citet{ste82} gives a general
discussion on the subject. One can distinguish between two main areas:
\\
(a) The analysis of effects due to time\hs independent gravitational
fields, such as Schwarzschild or Kerr spacetime, in an
astrophysical context \citep{MT,DT}, or \\
(b) scattering of EM fields off time\hs dependent gravitational fields
\citep{Cooperstock,Zeldovich,Gerlach,Denisov,Grishchuk-Polnarev,Demianski}.

The latter studies (b) have shown that there can be a coherent
interaction
between linear gravitational waves and EM fields, which would provide
an effective means of transferring energy from gravitaional to EM
degrees of freedom. It is for this reason that subsequent work focused
on using this mechanism as a way of detecting gravitational waves
\citep{Lupanov,Braginskietal,Grishchuk-Sazhin}.

Motivated by the discussion given in the references above, we return to
the time\hs dependent problem in order to further investigate a number
of effects which arise when a gravitational wave interacts with an EM
field,
this time with a plasma present.
In this letter we focus on a simple example, in order to demonstrate
the basic effect, and
in order to maximise clarity and make physical interpretations simple,
we will use the 1+3 orthonormal frame (ONF) formalism (see, e.g.,
\citet{cargese} and references therein). This has the advantage
of picking out a physical observer and splitting spacetime relative
to her.

The paper is organized as follows: In section \ref{sec:prel}
we briefly review the ONF formalism and rewrite Maxwells equations
using our ONF, introducing effective (gravity induced)
charge and current densities. Next, in section \ref{sec:excite},
we specify the metric to that of a weak one-dimesional gravitational
wave, and calculate the correpsonding (effective) current denstites.
Section \ref{sec:example} then considers the excitation of
an electromagnetic wave by a gravitational wave in a magnetized thin
one\hs component plasma. We deduce that the resulting electromagnetic
field amplitude is proportional to the gravitational wave amplitude,
the magnitude of the static magnetic field and the size
of the interaction region. In section \ref{sec:astro} we consider
the interaction region exterior to a binary pulsar. This consists
of a vacuum regime (swept out by the Poynting flux) and a
regime of interstellar matter. In order to simplify our analysis we
consider only regions in which the magnetic field is on average aligned
with the axis of rotation of the binary pulsar. Making an analog to
laboratory experiments we argue that efficient harmonic generation of
the EM waves may take place, and rough estimates suggest that these
harmonics might -- for favorable choices of parameters -- be
observable by the proposed Astronomical Low Frequency Array (ALFA)
\citep{Jones1}. Finally, our results are summarized and discussed
in section \ref{sec:summary}.

%%%%%%%%%%%%%%%%%%%%%%%%%%%%%%%%%%%%%
\section{Preliminaries}\label{sec:prel}
%%%%%%%%%%%%%%%%%%%%%%%%%%%%%%%%%%%%%
Suppose an observer moves with 4-velocity $u^a$ ($a = 0, ..., 3$), where
$u_au^a = 1$ (we put $c = 1$). This observer will measure the electric
and magnetic fields \citep{cargese}
\begin{equation}
  E_a \equiv F_{ab}u^b \ , \quad
  B_a \equiv {\onehalf}\,\epsilon_{abc}F^{bc} \ ,
\end{equation}
respectively, where $F_{ab}$ is the EM field tensor.
Here $\epsilon_{abc} \equiv u^d\eta_{dabc}$, where
$\eta_{abcd}$ is the totally skew 4-dimensional volume element, and
$\eta_{0123} = |\mathrm{det}(g_{ab})|^{1/2}$ (which will be just equal
to 1 in our ONF case). With these definitions, we can write the field
tensor as
\begin{equation}\label{eq:split}
  F^{ab} = u^aE^b - u^bE^a + \epsilon^{abc}B_c \ .
\end{equation}

We now introduce the ONF: $e_a$, $a = 0, ..., 4$, where $e_0 = u$,
so that, with the above splitting (\ref{eq:split}) of the field tensor,
Maxwell's equations $\nabla_bF^{ab} = \mu_0j^a$, $\nabla_{[a}F_{bc]} =
0$ read
\begin{mathletters}\label{eq:maxwell}
\begin{eqnarray}
  \boldsymbol{\nabla\cdot E} &=& \rho_{\!_E} +
\rho_\mathrm{m}/\varepsilon_0
  \label{eq:max1} \ , \\
  \boldsymbol{\nabla\cdot B} &=& \rho_{\!_B} \label{eq:max2}\ , \\
  e_0(\boldsymbol{E}) - \boldsymbol{\nabla\times B}
     &=& -\boldsymbol{j}_{\!_E}
         - \mu_0\boldsymbol{j}_\mathrm{m} \label{eq:max3}
     \ , \\
  e_0(\boldsymbol{B}) + \boldsymbol{\nabla\times E}
     &=& -\boldsymbol{j}_{\!_B}  \label{eq:max4} \ ,
\end{eqnarray}
\end{mathletters}
where the ``effective'' (gravity induced) charge densities and current
densities are given by
\begin{eqnarray}
  \rho_{\!_E} &\equiv& -\Gamma^{\alpha}\!_{\beta\alpha}E^{\beta}
  - \epsilon^{\alpha\beta\gamma}\Gamma^0\!_{\alpha\beta}B_{\gamma} \ ,
\label{eq:rhoe} \\
  \rho_{\!_B} &\equiv& -\Gamma^{\alpha}\!_{\beta\alpha}B^{\beta}
  + \epsilon^{\alpha\beta\gamma}\Gamma^0\!_{\alpha\beta}E_{\gamma} \ ,
\label{eq:rhob} \\
  \boldsymbol{j}_{\!_E} &\equiv& \left[ -(\Gamma^{\alpha}\!_{0\beta}
                       - \Gamma^{\alpha}\!_{\beta0})E^{\beta}
  + \Gamma^{\beta}\!_{0\beta}E^{\alpha}
  - \epsilon^{\alpha\beta\gamma}\left( \Gamma^0\!_{\beta0}B_{\gamma}
               + \Gamma^{\delta}\!_{\beta\gamma}B_{\delta} \right)
  \right]e_{\alpha} \ ,
\label{eq:je} \\
  \boldsymbol{j}_{\!_B} &\equiv& \left[ -(\Gamma^{\alpha}\!_{0\beta}
                       - \Gamma^{\alpha}\!_{\beta0})B^{\beta}
  + \Gamma^{\beta}\!_{0\beta}B^{\alpha}
  + \epsilon^{\alpha\beta\gamma}\left( \Gamma^0\!_{\beta0}E_{\gamma}
               + \Gamma^{\delta}\!_{\beta\gamma}E_{\delta} \right)
  \right]e_{\alpha}\ ,
\label{eq:jb}
\end{eqnarray}
while $\rho_\mathrm{m}$ and $\boldsymbol{j}_\mathrm{m}$ are the matter
charge and current densities, respectively.
Here $\Gamma^a\!_{bc}$ are the Ricci rotation coefficients, and we have
introduced the three\hs vector notation $\boldsymbol{E} \equiv
(E^{\alpha})
= (E^1, E^2, E^3)$ etc., and $\boldsymbol{\nabla} \equiv (e_1, e_2,
e_3)$.
The dot\hs\ and cross\hs products are defined in the usual Euclidian
way.

%%%%%%%%%%%%%%%%%%%%%%%%%%%%%%%%%%%%%%%%%%%%%%%%%%%%%%%%%%%%%
\section{EM wave excitation by weak gravitational radiation}%
\label{sec:excite}
%%%%%%%%%%%%%%%%%%%%%%%%%%%%%%%%%%%%%%%%%%%%%%%%%%%%%%%%%%%%%

To first order, the
solution to Einstein's vacuum equations can be written as (using the
transverse traceless gauge):
\begin{equation}\label{eq:gw}
  \mathrm{d}s^2 = -\mathrm{d}t^2 + (1+h)\mathrm{d}x^2
     + (1-h)\mathrm{d}y^2 + \mathrm{d}z^2 \ ,
\end{equation}
where $h$ is the deviation from flat space satisfying $|h| \ll 1$,
and we consider only the $+$\hs polarisation of the gravitational wave.
The deviation $h$ is a solution to the wave equation $\Box h = 0$,
i.e.\ $h = h(z-t)$ \citep{MTW}.
In order to clarify the physical intepretation and to simplify the
equations, we introduce a contravariant ONF
\begin{equation}
  e_0 = \partial_t \ , \ e_1 = \left(1 - {\onehalf}\,h\right)\partial_x
  \ , \ e_2 = \left(1 + {\onehalf}\,h\right)\partial_y
  \ , \ e_3 = \partial_z \label{eq:frame} \ ,
\end{equation}
such that the metric components become $\eta_{ab}
= \mathrm{diag}\,(-1,1,1,1)$.
Thus the local rest\hs space orthogonal to a synchronus observer takes
Euclidian form.

For the gravitational wave (\ref{eq:gw}), the charge densities
(\ref{eq:rhoe}) and (\ref{eq:rhob}) are zero. The current densities
(\ref{eq:je}) and (\ref{eq:jb}) have components
\begin{mathletters}\label{eq:current}
\begin{eqnarray}
  j^1_{\!_E} &=& -{\onehalf}\,(-E^1 + B^2)\,\dot{h} = j^2_{\!_B} \ ,
\label{eq:current1} \\
  j^2_{\!_E} &=& -{\onehalf}\,(E^2 + B^1)\,\dot{h} = -j^1_{\!_B} \ ,
\label{eq:current2}
\end{eqnarray}
\end{mathletters}
where $\dot{h} = \partial h/\partial z = -\partial h/\partial t$.

%%%%%%%%%%%%%%%%%%%%%%%%%%%%%%%%%%%%%%%%%%%%%%%%%%%%
\section{1\hs D example}\label{sec:example}
%%%%%%%%%%%%%%%%%%%%%%%%%%%%%%%%%%%%%%%%%%%%%%%%%%%%

Here we consider the effects of an incoming gravitational wave on a
static magnetic field in an inhomogeneous plasma, restricting our
attention to a 1\hs D case. Thus, we assume the presence of a external
static magnetic field ${\boldsymbol{\widehat B}}=\widehat{B}(z)e_1$,
and study gravitational wave excitation of small amplitude
plasma waves propagating in the $z$-direction. We take the
gravitational wave to be monochromatic, i.e.\ $h=\bar{h}\exp
[{\rm i}k(z-t)]$, and we let the frequency of the excited wave coincide
with the driver, that is, $E^{a} = \bar{E}^{a}(z)\exp(-{\rm i}\omega
t)$, where $\omega =k$, and similarly for all other quantities.
From Maxwells equations (\ref{eq:maxwell}), using $|\bar{B}(z)|,
|\bar{E}(z)|
\ll |\widehat{B}(z)|$ and equations (\ref{eq:current}),
we immediately obtain\footnote{We linearize such that
  terms quadratic in barred quantities are dropped.}
\begin{eqnarray}
  \left( \frac{\partial ^{2}}{\partial z^{2}}+k^{2} \right)\bar{E}%
  ^{2} + {\rm i}\omega \mu_0\bar{j}_{\rm m}^{2} &=&
  -\left( k^{2}{\widehat B} - \frac{1}{2}{\rm i}k\frac{d%
  \widehat{B}}{dz}\right) \bar{h}\exp ({\rm i}kz)
  \equiv S(z)\;, \label{eq:waveeq} \\
  \omega ^{2}\bar{E}^{3} + {\rm i}\omega\mu_0\bar{j}_{\rm m}^{3} &=&0\;.
  \label{eq:constraint}
\end{eqnarray}
In order to calculate the current, we assume the ions to be immobile
-- that is $\omega \gg \omega _{{\rm pi}}$, where $\omega _{{\rm pi}}$
is the ion plasma frequency. Using our ONF the equation of motion for
cold electrons can be written
\begin{mathletters}
\begin{equation}\label{eq:eqmot}
  mu^a\boldsymbol{v}_{;a} = q\left( v^{0}\boldsymbol{ E}
    + \boldsymbol{v\times B}\right)\;,
\end{equation}
where $v^a = (v^0, \boldsymbol{v})$ is the electron four\hs velocity,
satisfying $v^av_a = -1$, and $u^a$ is the observers four\hs velocity.
Linearizing the equation of motion (\ref{eq:eqmot}) around the
unperturbed plasma state (i.e.\ $\boldsymbol{E} = 0$,
$\boldsymbol{v} = 0$) and dropping all cross terms containting the
three\hs velocity and the rotation coefficients, we obtain
\begin{equation}\label{eq:eqmot2}
  -{\rm i}m\omega {\boldsymbol{\bar v}}= q\left( {\boldsymbol{\bar
        E}} + \boldsymbol{\bar{v}\times\widehat{
        B}}\right)\;.
\end{equation}
\end{mathletters}
Solving equation (\ref{eq:eqmot2}) for the electron velocity we obtain
\begin{mathletters}\label{eq:velocity}
\begin{eqnarray}
  \bar{v}^{2} &=& \frac{q}{m\omega }\left(
    {\rm i}\bar{E}^{2}+\frac{\omega_{\rm c}}{\omega
    }\bar{E}^{3}\right) \left( 1-\frac{\omega
    _{\rm c}^{2}}{\omega^{2}}\right)^{-1}\;, \label{eq:v2} \\
  \bar{v}^{3} &=& \frac{q}{m\omega }\left(
    {\rm i}\bar{E}^{3}-\frac{\omega_{\rm c}}{\omega
    }\bar{E}^{2}\right) \left( 1-\frac{\omega _{\rm c}^{2}}{\omega
    ^{2}}\right)^{-1}\;, \label{eq:v3}
\end{eqnarray}
\end{mathletters}
where $\omega_{\rm c}=q\widehat{B}/m$ is the electron cyclotron
frequency.
Using $\boldsymbol{j}_{\rm m} = qn_{0}\boldsymbol{v}$ where
$n_{0}$ is the unperturbed electron number density at rest, and
inserting equations (\ref{eq:velocity}) in
(\ref{eq:waveeq}) and (\ref{eq:constraint}), we obtain a
driven wave equation for the extra\hs ordinary
EM wave
\begin{equation}\label{eq:waveeq2}
  \left( \frac{\partial ^{2}}{\partial z^{2}} + k^{2} - \Delta
  k^{2}\right)\bar{E}^{2} = S(z)\;,
\end{equation}
where the plasma induced wavenumber mismatch $\Delta k$ is found to be
\begin{equation}\label{eq:mismatch}
  \Delta k^{2}=\frac{\omega _{\rm p}^{2}(\omega ^{2}-\omega
    _{\rm p}^{2})}{\omega^{2}-\omega _{\rm h}^{2}} \ .
\end{equation}
Here $\omega_{\rm p} = (n_{0}q^{2}/\varepsilon_{0}m)^{1/2}$
is the plasma frequency and $\omega_{\rm h} =
\left(\omega_{\rm p}^{2} + \omega_{\rm c}^{2}\right)^{1/2}$
is the upper hybrid frequency. Assuming that the
static field is localized in the region $0<z<a$,  the exact solution
to equation (\ref{eq:waveeq2}), when $\Delta k =$
constant,\footnote{Specifically this result holds in a vacuum region,
  and in this case the effect has been used as an argument for
  employing EM fields as gravitational wave detectors (see e.g.\
  \citet{Grishchuk-Polnarev} and references therein).} is
\begin{eqnarray}
  \bar{E}^{2}(z) &=& \exp \left[ {\rm i}k_{\rm loc}z\right]
  \int_{0}^{z}\exp \left[ -{\rm i}k_{\rm loc}z' \right]
  S(z')\mathrm{d}z' \nonumber \\
  && +\exp \left[ -{\rm i}k_{\rm loc}z\right]
  \int_{a}^{z}\exp \left[ {\rm i}k_{\rm loc}z'\right]
  S(z')\mathrm{d}z'\;, \label{eq:solution}
\end{eqnarray}
where the integration limits have been chosen so that there are no
waves coming into the excitation region. Here $k_{\rm loc} \equiv (k^2 -
\Delta k^2)^{1/2}$ is the local wave number of the electromagnetic
radiation. The solution (\ref{eq:solution}) also holds approximately
when both $|\Delta k| \ll |k|$ and $|d\Delta k/dz| \ll |k\Delta k|$.
We note that for
a large excitation region, $ka\gg 1$, EM waves propagating along the
positive $z$\hs axis can obtain much higher amplitudes than those in the
opposite direction, provided the mismatch is small, that is $\Delta k\ll
k$. In the next section we will consider the regime
\begin{equation}\label{eq:lowmismatch}
  \omega_{\rm p}^{2}/\omega_{\rm c} \ll \omega \lesssim
  \omega_{\rm p} \ll \omega_{\rm c} \ ,
\end{equation}
in which case the condition $\Delta k\ll k$ is fulfilled.
Focusing on this approximation, we may neglect the negative
propagating wave all together,
and write the solution in the outgoing region $z>a$ as
\begin{equation}
  \bar{E}^{2}(z,t)=E_{{\rm out}}\exp [{\rm i}(k_{\rm out}z - \omega
t)]\;,
\end{equation}
where
\begin{mathletters}
\begin{equation}\label{eq:Eampl}
  E_{{\rm out}}=-\frac{{\rm i}k\bar{h}}{2}\int_{0}^{a}\widehat{B}(z)\exp
  \left( \frac{{\rm i}\Delta k^2z}{2k} \right)\mathrm{d}z
\end{equation}
and $k_{\rm out} = k_{\rm loc}(z = a)$.\footnote{If we insist on a
  sharply defined excitation region $0 < z < a$, $k_{\rm out}$
  typically becomes imaginary, reflecting the fact that waves with
  $\omega < \omega_{\rm p}$ does not propagate in an unmagnetized one\hs
  component plasma.}
This shows that the effective size of the ''gravitational
EM wave transmitter'' is determined by either the extension of the
magnetic field, provided the direction of the static magnetic field does
not oscillate, or the mismatch distance $L=2k/\pi\Delta k^2$, whichever
is
smaller. This is a consequence of the fact that extra\hs ordinary mode
and
gravitational waves satisfy nearly the same dispersion relation in the
regime of consideration, which tends to make linear wave interaction
coherent over large distances.

%%%%%%%%%%%%%%%%%%%%%%%%%%%%%%%%%%%%%%%%%%%%%%%%%%%%%%%%%%%%%%%%%%%%%%%
\section{Applications to astrophysics}\label{sec:astro}
%%%%%%%%%%%%%%%%%%%%%%%%%%%%%%%%%%%%%%%%%%%%%%%%%%%%%%%%%%%%%%%%%%%%%%%

It is an interesting question whether it is possible to detect the
generated
EM waves during realistic conditions. An immediate problem is that for
most binary astrophysical sources, such as binary neutron stars
\citep{Rasio-Shapiro}, $\omega/2\pi \lesssim 10^{3}{\rm Hz}$, whereas
the
plasma density for typical interstellar matter correspond to
$\omega _{{\rm p}}/2\pi \gtrsim 10^{3}{\rm Hz}$.
Although an external magnetic field may decrease the cut-off frequency
significantly below $\omega _{{\rm p}}$ for EM waves propagating {\em \
perpendicular} to the magnetic field, we cannot rely on that effect
for EM wave propagation over interstellar distances. Furthermore,
observation from the ground are inaccessible for frequencises roughly
below 10 MHz, due to the ionospheric cut\hs off. However, as we will
demonstrate below, it is too early to conclude that gravitationally
induced EM waves cannot be detected.

Let us consider a binary system consisting of two identical
pulsars,\footnote{One might be hesitant to apply our results derived
  for a one\hs component plasma, to the electron--positron plasma
  that exists in the vicinity of pulsars. For our case, however,
  the presence of the plasma is not very important in the EM excitation
  region, due to the strong magnetization which keeps down the
  mobility, but instead the plasma affects the EM wave propagation
  further away from the binary pulsar, as to be discussed below.
  For a discussion of EM excitations by gravitational waves in a
  multi\hs component plasma, see \citet{Ignatev}.}
with individual masses $M_{\sun}$ (for a general review of
the physical properties of binary pulsars, see
\citet{Phinney-Kulkarni}).
This leads to a Schwarzschild radius $R_{{\rm S}} =
2GM_{\sun}/c^2 \approx 3$ km. Furthermore, we assume that the
separation distance
between the two neutron stars is $d=20R_{{\rm S}}$. Using a Newtonian
approximation, we find that the gravitational wave frequency is
$\omega \approx 2\times 10^{3}$ rad/s.
This corresponds to a situation fairly close to collapse. Using the
quadrupole radiation formula (see, e.g., \cite{ste82}), assuming two
Newtonian point masses, the expected
time before merging is of the order of $10^{-2}\ {\rm
  s}$.\footnote{During the last part of the inspiral the radiation is
  far from monochromatic. We do not believe that this will have any
  significant effect on the physics discussed.}
The unperturbed magnetic field is taken as $B(r) = B_{{\rm surf}}r_{{\rm
puls%
}}^{3}/r^{3}$, where the surface magnetic field
$B_{{\rm surf}}$ is of the order $10^{6} - 10^{10}$ T (where the upper
limit
refers to magnetars), and $r_{{\rm puls}}$ is the radius of the pulsar.
Thus
the fall\hs off of the magnetic field is that of a dipole.

Next we want to estimate the maximum electric field from equation
(\ref{eq:Eampl}). To get a magnetic field that is essentially
static in the frame of reference where the binary system is rotating,
we must exclude the region closest to the pulsar surface from the
region of interaction. Assuming that the
electromagnetic excitation effectively starts at a distance $r=60R_{{\rm
S}}$
from the center of mass of the system, using the estimates presented
above,
together with condition (\ref{eq:lowmismatch}), we find from equation
(\ref{eq:Eampl}) that
\begin{equation}
E_{\max }\sim 1.5\times 10^{6}\ \bar{h}_{{\rm int}}B_{{\rm surf}}\ {\rm
V/m}
\ , \label{eq:Emax}
\end{equation}
where we have denoted the gravitational wave amplitude
at the beginning of the interaction
region ($r=60R_{{\rm S}}$) by $\bar{h}_{{\rm int}}$. Here we have used
that
the effective interaction distance is determined by a fall off in the
interaction efficiency together with a spherical attenuation of the
electromagnetic wave, rather than a decoherence of the modes as
discussed in
the analytical calculations. Using\footnote{%
Using a Newtonian approximation with $d = \alpha R_{{\rm S}}$,
$r = \beta R_{{\rm S}}$, we find that $\bar{h} \sim
(2\alpha\beta)^{-1}$.}
$\bar{h}_{{\rm int}}\sim 0.001$
and $B_{{\rm surf}}\sim 10^8$ T, equation (\ref{eq:Emax}) gives
\begin{equation}
E_{{\rm max}}\sim 50\ {\rm MV/m}\ ,
\end{equation}
\end{mathletters}
at a distance $r_{\max }\approx 120R_{{\rm S}}$. In this region, the
induced
electron velocity is kept non-relativistic -- to a good approximation --
by
the external magnetic field, since the electrical and magnetic fields
are
perpendicular, and we have $v\approx E_{{\rm max}}/B(r_{\max })\approx
6\times 10^{5}$ m/s. Using an $1/r$ spherical decay of the EM field for
$r>$ $r_{\max }$, the electron velocity can also be estimated at large
distances. In a distant region where the magnetic field is assumed to
have
decreased such that $\omega _{\rm c}\lesssim \omega $,
the electron velocity is
of the order $v(r)\sim qE_{{\rm max}}r_{\rm max }/(m\omega r)$,
provided
the ultrarelativistic regime is not reached. The radius $r_{{\rm rel}}$
from
the source where the electron \emph{quiver velocity} is moderately
relativistic (which we define to be $v=10^{8}$ m/s) can then be solved
for.
We obtain
\begin{equation}
r_{{\rm rel}}\sim 7.5\times 10^9 \ {\rm km} \sim 500\ {\rm au}\ .
\end{equation}
Since the magnetic field from the binary system is weak at
this distance (i.e. the condition $\omega _{\rm c}\lesssim \omega $ is
fulfilled), we deduce that the electron velocity must become
relativistic at
some distance $r$ satisfying $r_{\rm max }$ $\ll r\ll r_{{\rm rel}}$ --
the
precise value depend on extension of the magnetic field.\footnote{%
It should be noted that the $1/r^3$ behaviour of the magnetic field
holds to
the light cylinder, outside of which the magnetic field falls off
as $1/r$ \citep{Rees}.}

When the electron velocities approaches the speed of light the excited
EM wave becomes highly nonlinear, as the ponderomotive and relativistic
nonlinearities in the momentum equation becomes comparable to the linear
terms \citep{Shukla}. Several nonlinear effects may take place, such
as parametric excitation of other plasma waves (exciting lower
frequencies) and harmonic generation of the original mode.
A detailed study of the evolution of the EM waves is by far beyond
the scope of our investigation. However, we note that the degree
of inhomogeneity of the background medium may be crucial for the
outcome. From now on we assume that the nonlinear regime is reached
comparatively close to the binary source, where the plasma medium can be
characterized as strongly imhomogeneous. This means that the
background parameters $\omega _{\rm c}$ and $\omega _{{\rm p}}$ are
assumed to change significantly during a single wavelength of the
EM wave.  Strong inhomogeneities generally
tend to supress parametric excitations \citep{Liu}, and thus we expect
harmonic generation to be the more important process in our case.
To get a feeling for the possible magnitude of this effect,
we can compare with laser experiments
\citep{Carman-Forslund-Kindel,Carman-Rhodes-Benjamin},
where the relativistic velocities induced in an inhomogeneous plasma
leads to repeated harmonic generation by
a factor of order 50 of the original laser frequency,
and these harmonics have a comparable amplitude to that of the
original frequency (see \citet{Bezzerides-Jones-Forslund} and
\citet{Grebogi-Tripathi-Chen} for a theoretical explanation of this
effect). For the parameters chosen in our example, the electron
velocity can be as large as in the experiments presented by
\citet{Carman-Forslund-Kindel} and \citet{Carman-Rhodes-Benjamin}.
The inhomogeneity parameter $1/kL$, where $L$ is the inhomogeneity scale
length, is also of comparable magnitude to the same experiments,
and thus it is not unrealistic to assume that a significant part of the
excited EM energy can be converted to higher harmonics with a harmonic
number between 40 and 50 of the original frequency, which correspond
to a EM wave frequency in the long wavelength radio regime, but well
above the typical
cut-off frequency of the interstellar plasma. Such EM waves can
propagate over
astrophysical distances and suffer no other significant damping than
spherical
attenuation.

Detection of such low frequecy signals cannot be done on earth,
due to absorption and reflection of the ionospheric plasma.
Thus it is evident that no existing radio
telescopes can measure the electromagnetic emissions discussed
above, but well developed proposals like the Astronomical Low
Frequency Array (ALFA) is expected to measure
electromagnetic signals in the interval 30 kHz to 30 Mhz
\citep{Jones1}. Assuming that 10\% of the energy gets transferred
to harmonic numbers with a high enough
frequency for detection (roughly harmonic number $n\gtrsim 60
$), and estimating the minimum detection level (due to noise
originating from the solar system) to 1000 Jy \citep{Jones2},
the maximum distance of observation becomes $\sim 3\times 10^5\ {\rm
ly}$
if we assume that the radio signal is evenly distributed in the
frequency range 30--40 kHz.
Thus we deduce that the source considered above
can be observed by a system such as ALFA provided it is located
to our own galaxy.\footnote{Furthermore, we expect
  (at least for neutron stars with stiff
  equations of state \citep{Rasio-Shapiro}) that at the final stage
  of the inspiral process, the frequency and amplitude of the
  gravitational radiation will increase, making the process
  more effective over a short time\hs span. Also a larger distance of
  observation can be achieved if we assume that the harmonic structure
  of the EM wave is not destroyed during its generation.}
The theoretical estimates of the event rates of collapsing
binaries in our galaxy are of the order of $10^{-4}\ {\rm yr}^{-1}$
\citep{Tutukov-Yungelson,Yamaoka-etal} (for a review, see
\cite{Schutz}).

%%%%%%%%%%%%%%%%%%%%%%%%%%%%%%%%%%%%%%%%%
\section{Summary and discussion}\label{sec:summary}
%%%%%%%%%%%%%%%%%%%%%%%%%%%%%%%%%%%%%%%%%

In this letter we have investigated the interactions between weak
gravitational waves and EM fields, using the 1+3 orthonormal frame
formalism. In this way we were able to give a clear presentation of
the result that an EM wave can be generated as a result of
the interaction of a gravitational wave with a static isolated
magnetic field. Furthermore, in the presence of a plasma we argued that
higher harmonics of the original EM wave may be generated, and an
astrophysical example, consisting of a binary system, has been
considered.

Compact binaries, although not too common, are interesting sources of
gravitational radiation,
since the theoretical predictions of their spectral charachteristics
are believed to be well understood. It is therefore an interesting
question whether such charachteristics would be detectable through the
radio wave emissions described above. In our case, the harmonic
generation mechanism must be analyzed in more detail in order for
useful spectral predictions to be made. If this was to be
accomplished, it would shed light on the
possible nature of the radio source, and thus would provide us with an
indirect observational method for gravitational radiation.

In addition to binary systems, there are of course other simultaneous
sources of magnetic fields and gravitational radiation, e.g.\ quaking
neutron stars. One of the more extreme events of this kind would be
the quaking magnetar, with a magnetic field of the order of $10^{10}\
{\rm T}$. The gravitational wave frequency of such systems may also be
up to one
order of magnitude higher than in the example considered above (at
least for the gravitational wave {\it w}-modes
\citep{Kokkotas-Apostolatos-Andersson}). However, the physics involved
in generating the gravitational waves tends to be more complicated,
and thus the corresponding estimates of the amplitudes and frequencies
are more insecure
\citep{Nisse,Andersson-Kokkotas,Kokkotas-Apostolatos-Andersson}. It
may still be of interest for further studies, since the large magnetic
fields of magnetars could compensate for the expected lower amplitude of
the gravitational waves.

It is of interest to consider possible effects due to strong
gravitational field deviations from Minkowski spacetime.
According to \citet{DT} all frequencies, e.g.\ the plasma\hs\ and
cyclotron frequencies, are redshifted by the lapse function
in a spherically symmetric spacetime.
Since the important quantity in the above calculations are the
{\em quotient} of certain frequencies, the effect will persist, if
we assume that the present situation can be approximated as spherical
symmetric. Of course, the frequency of the emitted radiation will be
redshifted ($\omega_{\text{emmision}}/\omega_{\text{reception}} = 1 +
z$), although $z$ will be less than 2\% for the interaction
distance $60R_{\rm S}$ used in the above example.

In light of these results we will discuss the possible
cosmological implications in a forthcoming paper
\citep{Marklund-Dunsby-Brodin}, using the 1+3 covariant approach and
tetrad methods which in a clear unified way deals with such
problems.  For example, we speculate that this may lead to an imprint
on the Cosmic Backround Radiation (CBR) spectrum through the
interaction between primordial gravitational waves and magnetic
fields of cosmological origin. Therefore CBR measurements could in
principle be used to constrain cosmological magnetic fields.

%%%%%%%%%%%%%%%%%%%%%%%%
\acknowledgments
%%%%%%%%%%%%%%%%%%%%%%%%
  M.\ M.\ was supported by the Royal Swedish Academy of
  Sciences. P.\ K.\ S.\ D.\ was supported by the NRC (South Africa) and
  the URC (University of Cape Town). The authors would like to thank
  the referee for raising the question of core/crust quakes in
  magnetars.

\end{document}